\documentclass[reprint,prc,aps,superscriptaddress,floatfix,showpacs,showkeys,altaffilletter]{revtex4-2}

\newcommand{\nuc}[2]{$^{\text{#1}}$#2}
\newcommand{\rme}[3]{\langle{#1} \lVert{#2} \rVert{#3} \rangle}

\usepackage{comment}
\usepackage[pdftex]{graphics}      
\usepackage{graphicx}                  
\usepackage[rightcaption]{sidecap}
\usepackage{epsfig}
\usepackage{bm}
\usepackage{verbatim}
\usepackage{color}
\usepackage[percent]{overpic}
\usepackage{multirow}
\usepackage{longtable}
\usepackage{tablefootnote}
\usepackage{footnote}
\usepackage{dcolumn}
\usepackage{tabularx}
\usepackage{tablefootnote}
\usepackage{array}
\usepackage{epstopdf}
\usepackage{hyperref}
\usepackage{amsmath}
\usepackage[T1]{fontenc}
\usepackage{tikz}

\hypersetup{
	colorlinks   = true,
	citecolor    = blue,
	urlcolor     = blue,
	linkcolor   = blue,
}

\begin{document}


\title{Coulomb excitation of \nuc{\bf{124}}{Te}: Emerging collectivity and persisting seniority structure in the \boldmath{$6_1^+$} level}



\author{M.~Reece}
\affiliation{Department of Nuclear Physics and Accelerator Applications, Research School of Physics, The Australian National University, Canberra, ACT 2601, Australia}

\author{B.~J.~Coombes}
\affiliation{Department of Nuclear Physics and Accelerator Applications, Research School of Physics, The Australian National University, Canberra, ACT 2601, Australia}

\author{A.~J.~Mitchell}
\email[Email: ]{aj.mitchell@anu.edu.au}
\affiliation{Department of Nuclear Physics and Accelerator Applications, Research School of Physics, The Australian National University, Canberra, ACT 2601, Australia}

\author{A.~E.~Stuchbery}
\affiliation{Department of Nuclear Physics and Accelerator Applications, Research School of Physics, The Australian National University, Canberra, ACT 2601, Australia}

\author{G.~J.~Lane}
\affiliation{Department of Nuclear Physics and Accelerator Applications, Research School of Physics, The Australian National University, Canberra, ACT 2601, Australia}

\author{A.~Gargano}
\affiliation{Istituto Nazionale di Fisica Nucleare, Complesso Universitario di Monte S. Angelo, Via Cintia, I-80126 Napoli, Italy}

\author{V.~U.~Bashu}
\affiliation{Department of Nuclear Physics and Accelerator Applications, Research School of Physics, The Australian National University, Canberra, ACT 2601, Australia}
\affiliation{ARC Centre of Excellence for Dark Matter Particle Physics, Australia}

\author{L.~J.~Bignell}
\affiliation{Department of Nuclear Physics and Accelerator Applications, Research School of Physics, The Australian National University, Canberra, ACT 2601, Australia}
\affiliation{ARC Centre of Excellence for Dark Matter Particle Physics, Australia}

\author{C.~Gautam}
\affiliation{Department of Engineering Physics, Air Force Institute of Technology WPAFB, OH 45433, USA}

\author{L.~J.~McKie}
\affiliation{Department of Nuclear Physics and Accelerator Applications, Research School of Physics, The Australian National University, Canberra, ACT 2601, Australia}
\affiliation{ARC Centre of Excellence for Dark Matter Particle Physics, Australia}

\author{N.~J.~Spinks}
\affiliation{Department of Nuclear Physics and Accelerator Applications, Research School of Physics, The Australian National University, Canberra, ACT 2601, Australia}
\affiliation{ARC Centre of Excellence for Dark Matter Particle Physics, Australia}

\author{J.~A.~Woodside}
\affiliation{Department of Nuclear Physics and Accelerator Applications, Research School of Physics, The Australian National University, Canberra, ACT 2601, Australia}

\date{\today}


\begin{abstract}
The low-lying energy spectra of even-even tellurium isotopes near midshell have long been interpreted as `textbook' examples of vibrational collective motion. However, in many cases electric-quadrupole observables, which are a particularly sensitive probe of collectivity, remain undetermined. Coulomb-excitation measurements were performed to measure transition strengths connecting the ground and low-excitation states in \nuc{124}{Te}. This isotope lies at a transitional point between collective structure near the neutron midshell and seniority structures near the $N=82$ shell. A transition strength, $B(E2; 6_1^+ \to 4_1^+)$, of 27(9)~W.u. was measured for the $6^+_1\rightarrow4^+_1$ transition for the first time in this nucleus; this value is significantly below that expected for a spherical vibrator, as well as other collective models. We examine the transition strengths in \nuc{124}{Te} and its neighbors by comparison with large-basis shell-model calculations and by comparison with General Collective Model (GCM) fits. A GCM description of \nuc{120}{Te} agrees with experimental $E2$ transition strengths, but no comparable description of \nuc{124}{Te} is possible with the GCM. In contrast, there is remarkably good agreement between the $B(E2; 6_1^+ \to 4_1^+)$ values and shell-model calculations for \nuc{124--134}{Te}. It appears that, despite approaching midshell, \nuc{124}{Te} retains a seniority structure for the $6^+_1$ level, i.e.\ a significant $\pi 0g_{7/2}^2$ contribution. This persistence of the shell structure at the $6^+_1$ state is in contrast to the $B(E2)$ values of the lower-excitation $2^+_1$ and $4^+_1$ states in \nuc{124}{Te}, and neighboring \nuc{120}{Te} and \nuc{122}{Te}, for which the collectivity becomes enhanced as more neutrons are removed from $N=82$.
\end{abstract}

\maketitle

\section{INTRODUCTION}\label{sec:introduction}
Atomic nuclei immediately above and below the $Z = 50$ shell closure have long been described as `textbook' examples of spherical, vibrational nuclei~\cite{Rowe_2010}.
For those isotopes near the neutron midshell, patterns emerge in the spacing of the low-lying level energies and their associated spins, which show close agreement with those predicted by the vibrational model~\cite{Bohr_1953,KERN_1995}.
The historic view is that single-particle behavior gives way to collective, coherent motion of nucleons via a region of spherical vibrators between magic numbers and deformed nuclei at midshell.
However, a growing body of evidence from other spectroscopic observables, such as transition strengths and quadrupole moments, has initiated a deeper consideration of the assignments made to so-called vibrational states in this region~\cite{Garrett_2010,Garrett_2018,Garrett_2019}.

Tellurium ($Z = 52$) lies two protons beyond the shell closure at $Z = 50$.
Low-energy excitations of the neutron-rich Te isotopes near the $N = 82$ closed shell are well described by the shell model and hence can be considered as largely single-particle in nature~\cite{Allmond_2014,Kumar_2022,Wang_2020,Stuchbery_2013}.
As neutrons are removed along the isotope chain, collective properties are expected to become enhanced toward the midshell at $N=66$, i.e.\ \nuc{118}{Te}~\cite{118Te}.
These nuclei, with vibrational-like level sequences, therefore mirror the Cd isotopes that have been more thoroughly measured and discussed~\cite{Garrett_2010,Garrett_2018,Garrett_2019}.

\begin{figure*}[t]
	\includegraphics[width = 17.8 cm]{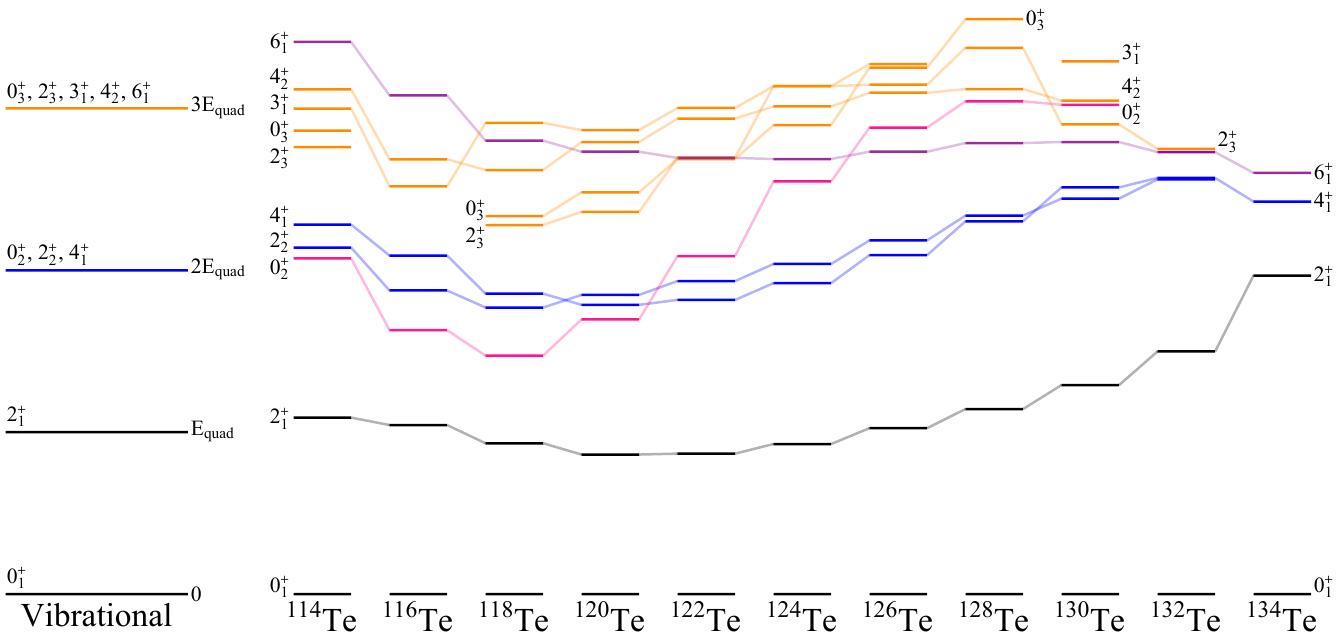}
	\caption{
		Systematic evolution of low-lying excited states in even-even tellurium isotopes from below the midshell ($N = 66, A = 118$) to the neutron shell closure ($N = 82, A = 134$).
		Degenerate groupings of states predicted to occur from one-, two-, and three-phonon excitations, with energies scaled relative to the $2^+_1$ level in \nuc{124}{Te}, are included on the left for reference.
		Experimental data (level energies and spin-parity assignments) are taken from Refs.~\cite{114Te,116Te,118Te,120Te,122Te,124Te,126Te,128Te,130Te,132Te,134Te}.\label{fig:1}
	}
\end{figure*}

The systematic evolution of low-lying states observed in even-even Te isotopes, from the midshell to $N = 82$, is summarized in Fig.~\ref{fig:1}.
The empirical data are shown along with a representation (on the left) of the collections of degenerate one-, two-, and three-phonon quadrupole excitations predicted by the vibrational model.
The lighter isotopes exhibit states with the spins and parities expected by the vibrational model, and these appear at energies near those expected for the multiphonon excitations.
However, several features point toward more complex underlying behavior that cannot be accounted for by the vibrational model alone.

A parabolic trajectory observed for the $0^+_2$ state energy, forming a minimum at \nuc{118}{Te}, was an early indicator of deficiencies in the vibrational description of Te isotopes~\cite{RIKOVSKA_1989}, resulting in ongoing discourse over the broader impact of shape coexistence in the region~\cite{Garrett_2022,Sabri_2016,Heyde_2011,Bonatsos_2022}.
This parabolic behavior is characteristic of shape-coexisting intruder states, and a similar pattern is also observed in the neighboring Sn and Cd isotopes~\cite{Garrett_2022}.
Other work has proposed that the raised $0^+_2$ energy in \nuc{124}{Te} indicates that a transition from vibrational to rotational behavior has occurred in this nucleus~\cite{Robinson_1983}.
In either case, the removal of the $0^+_2$ state from the assumed two-phonon triplet makes a vibrational interpretation untenable, as the colocation of the $0^+_2$ state with the $4^+_1$ and $2^+_2$ states in the two-phonon triplet is foundational to the vibrator model~\cite{Scharff_Goldhaber_1955,KERN_1995}.

The behavior of \nuc{124}{Te} has elicited extensive discussion and investigation as, moving from midshell toward $N = 82$, it is the lightest isotope to deviate noticeably from the energy pattern of the vibrational model; see Fig.~\ref{fig:1}.
Based on interacting boson model and dynamic pairing-plus-quadrupole model calculations, Gupta~\cite{Gupta_2023} suggested that a weakly deformed vibrational model is sufficient to explain the $B(E2)$ values and level energies of \nuc{118--124}{Te}.
However, these calculations were unable to effectively reproduce the level properties in \nuc{124}{Te} for states above the notional second-phonon triplet.
Beyond the vibrational picture, \nuc{124}{Te} has been proposed as a candidate for the E(5) critical-point symmetry between spherical vibration and triaxial rotation~\cite{Clark_2004,Arias_2001,Zhang_2019,Ghita_2008}.
This model has recently been explored through lifetime measurements~\cite{Hicks_2017} and high-spin spectroscopy~\cite{Tiwary_2019}.

Hicks \textit{et al.}~\cite{Hicks_2017} compare transition strengths in \nuc{124}{Te}, determined from measured excited-state lifetimes, with E(5) critical-point symmetry model predictions~\cite{Clark_2004}.
While many of the decay properties of \nuc{124}{Te} are described well by the E(5) symmetry, the decay of low-lying $0^+$ states deviates from the E(5) prediction, which better describes these decays in \nuc{122}{Te} and \nuc{126}{Te}.
Tiwary \textit{et al.}~\cite{Tiwary_2019} present evidence for $\gamma$-soft behavior on the basis of odd-even energy staggering of rotational-band members, although this inference is based on a limited range of spin values and unconfirmed assignment of odd-spin members of the so-called `$\gamma$ band'.

A recent theoretical investigation suggested the emergence of triaxial shapes around $A=124$ as a transition from oblate ground-state shapes in $114 \leq A \leq 124$~\cite{Sharma_2019}.
However, the potential energy surfaces in all cases show soft, weakly deformed minima.
Comparisons with experimental data were restricted to ground-state observables.

Coulomb-excitation measurements by Saxena \textit{et al.}~\cite{Saxena_2014} offer insights into the excited-state structure through the measurement of transition strengths ($B(E2)$ values) across a range of neighboring isotopes, \nuc{120,122,124}{Te}.
In their study, measured $B(E2)$ values were found to be consistent with a triaxial-rotation description for all three isotopes.
Further measurements on \nuc{120}{Te} by Saxena \textit{et al.}~\cite{Saxena_2019} allowed the evaluation of quadrupole invariants for the ground state and first-excited state.
These data affirmed a quasi-rotational view, and the observation that simple schematic models are insufficient; rotational and vibrational degrees of freedom must be considered together.
These conclusions, however, are limited by the absence of $B(E2)$ measurements involving states above the notional second-phonon triplet.

Beyond collective deformations, some authors have also suggested persisting aspects of the seniority scheme from near $N = 82$ as a possible explanation of the excited states in \nuc{124}{Te}.
Kerek \textit{et al.}~\cite{Kerek_1971} observed that while the energy ratio of the $4^+_1$ and $2^+_1$ states is close to the vibrational limit for \nuc{114--130}{Te}, the energy of the $6^+_1$ state remains flat with increasing neutron number.
This near-constant trend is at odds with the behavior of the other low-lying states, which tend to increase in energy from midshell toward the $N = 82$ shell closure.
Lee \textit{et al.}~\cite{Lee_1991} proposed that the $2^+_1$ states in tellurium exhibit vibrational behavior, whereas the $6^+_1$ states arise from two-proton coupling, and the $4^+_1$ states are a mixture of both.
At the $N = 82$ neutron shell closure (\nuc{134}{Te}), the low-lying states and transitions are described well by $\pi{(g_{7/2})}^2$ coupling~\cite{Stuchbery_2013}, and the coupling of two protons to a quadrupole vibrator can reproduce the $6^+_1$ energy in \nuc{124}{Te}~\cite{Lopac_1970}.
Shell-model calculations across the isotope chain have also been successful in reproducing the reduction in the energy gap between the $4^+_1$ and $6^+_1$ states approaching $N = 82$~\cite{Qi_2016}.
However, in an increasing number of investigations of developing collectivity, measurements of additional observables, particularly electric-quadrupole transition strengths, have illuminated different behavior than is inferred from energy systematics alone (see, for example, Refs.~\cite{Gerathy_2021,Garrett_2010,Saxena_2014}).
In \nuc{124}{Te}, the absence of $B(E2)$ and $g$-factor measurements has prevented a thorough assessment of seniority-like behavior above the first few excited states.

To date, no reliable measurement of the $B(E2;6^+_1 \to 4^+_1)$ value in \nuc{124}{Te} has been made.
Additionally, there are discrepancies between existing $B(E2)$ values determined from lifetime measurements~\cite{Hicks_2017} and Coulomb-excitation measurements~\cite{Saxena_2014} for transitions between several other low-lying states.

An experimental setup to perform multistep Coulomb-excitation measurements with particle-$\gamma$ coincidence spectroscopy has recently been established at the Heavy Ion Accelerator Facility at the Australian National University.
Here, we present results of a Coulomb-excitation measurement on \nuc{124}{Te}, the first dedicated experiment to use the new setup.
We have extracted $E2$ matrix elements from the measured transition yields using the semiclassical Coulomb-excitation code, \textsc{Gosia}~\cite{Cline_2012}, including that of the $6^+_1 \to 4^+_1$ transition.
The experimental results are compared to shell-model calculations and the general collective model (GCM).
These two extreme models, one single-particle based and the other wholly collective, enable a broad discussion on the structure of \nuc{124}{Te}, the persistence of seniority structure, and the emergence of collectivity.

\section{EXPERIMENT DETAILS}\label{sec:experiment}
The experiments were performed at the Heavy Ion Accelerator Facility at the Australian National University.
Beams of $185.5$-MeV \nuc{58}{Ni} and $45$-MeV \nuc{16}{O} were delivered by the 14UD Pelletron accelerator~\cite{OPHEL_1974}, with intensities of 0.7~particle~nA and 0.5~particle~nA, respectively.
These beams were incident on a 300-$\mu$g/cm$^2$ thick target of \nuc{124}{Te} evaporated onto a 20-$\mu$g/cm$^2$ thick backing of \nuc{12}{C}, similar to the procedures outlined in Ref.~\cite{Stolarz_1999}.
The `safe' energies for Coulomb excitation of these beam-target combinations are 191.5 MeV for \nuc{58}{Ni} and 47 MeV for \nuc{16}{O}~\cite{Cline_1986}.
The isotopic enrichment of \nuc{124}{Te} in the target material was $90\%$; the isotopic assay is listed in Table~\ref{tab:tgtassay}.

\begin{table}[h]
	\begin{center}
		\caption{\label{tab:tgtassay} Percent abundance of isotopes in the target.}
		\begin{tabular*}{0.48\textwidth}{@{\extracolsep{\fill}}cccccccc}
			\toprule \\[-0.3cm]
			\nuc{120}{Te} & \nuc{122}{Te} & \nuc{123}{Te} & \nuc{124}{Te} & \nuc{125}{Te} & \nuc{126}{Te} & \nuc{128}{Te} & \nuc{130}{Te} \\
			$<0.006$   & $0.09$ & $0.08$ & $90.0(4)$ & $3.45$ & $4.05$ &  $1.5$ & $0.83$\\
			\hline
			\hline
		\end{tabular*}
	\end{center}
\end{table}

\begin{figure}[t]
	\includegraphics[width = 8.6 cm]{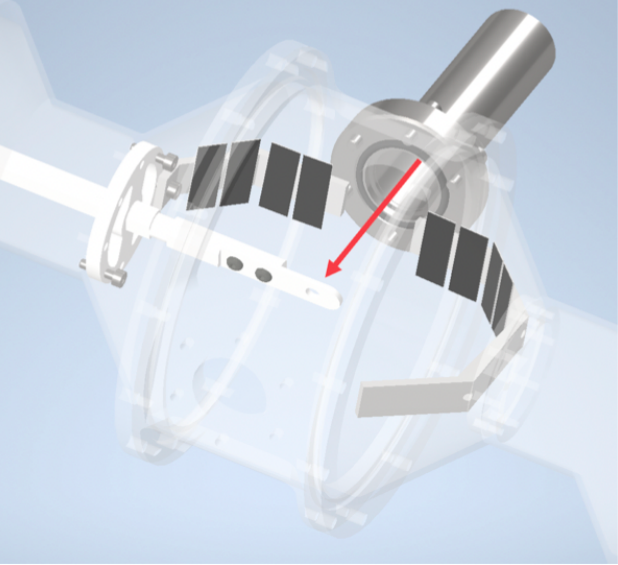}
	\caption{
		A CAD model of the new CAESAR target chamber showing the positions of the silicon photodiodes (dark rectangles) used in this work.
		The red arrow indicates the beam direction and the target position is at the end of the arrowhead.
		The photodiodes from left to right in this diagram are designated by the labels `$a$-$h$' in the text.\label{fig:2}
	}
\end{figure}

States excited in \nuc{124}{Te} were studied via particle-$\gamma$ coincidence spectroscopy using the CAESAR array of eight Compton-suppressed High-Purity Germanium (HPGe) detectors~\cite{Dracoulis_1989}.
To facilitate Coulomb-excitation measurements with CAESAR, a new target chamber and a particle detection array consisting of eight rectangular silicon photodiodes with active areas of $25.17 \times 9.25$~mm$^2$ were developed.
The detectors were mounted on a pair of aluminum support frames; the frames are versatile and can be fastened at eight different angles on each side of the target chamber.
For the current experiments, the photodiodes were positioned symmetrically about the beam axis at backward angles in the horizontal plane.
A CAD model of this arrangement is shown in Fig.~\ref{fig:2}.

Two near-identical implementations were used.
In the first case, used for the \nuc{16}{O} beam measurements, the photodiodes were wrapped in a layer of aluminized Mylar film 6-$\mu$m thick, to create a light-tight seal.
The photodiodes in the second implementation, used for the \nuc{58}{Ni} beam measurements, were wrapped in layers of 2-$\mu$m thick Mylar film and 0.74-$\mu$m thick aluminum foil, again to create a light-tight seal.
Polar angles and solid angles for both implementations are given in Table~\ref{tab:2}.

\begin{table}[t]
	\begin{center}
		\caption{
			Positions of the silicon photodiode particle detectors used in this work.
			The $z$ axis is defined to be the beam direction, $\theta$ is the polar angle and $\phi$ is the azimuthal angle, with $\phi = 0^{\circ}$ defined as vertically upwards.
            The photodiodes were replaced between the measurements with \nuc{16}{O} and \nuc{58}{Ni} beams.
			Detector $c$ was faulty and removed from the data stream during the \nuc{58}{Ni} run.
			Angles are given in degrees, 
   solid angles, $\Omega$, are given in msr.\label{tab:2}
		}
		\renewcommand{\arraystretch}{1.5}
		\begin{tabular*}{0.48\textwidth}{@{\extracolsep{\fill}}crrrrrr}
			\toprule
			&& \multicolumn{2}{c}{\nuc{16}{O} beam} & \multicolumn{2}{c}{\nuc{58}{Ni} beam}\\ \cline{3-4} \cline{5-6}
			Photodiode &$\phi^{\circ}$ & $\theta^{\circ}$ & $\Omega$~(msr) & $\theta^{\circ}$ & $\Omega$~(msr)  \\
			\hline
			$a$ & 90  & 122 & 62  & 121 & 60 \\
			$b$ & 90  & 133 & 64  & 131 & 65 \\
			$c$ & 90  & 144 & 64  & ---   & ---  \\
			$d$ & 90  & 154 & 98  & 153 & 95 \\
			$e$ & 270 & 151 & 106 & 154 & 101\\
			$f$ & 270 & 141 & 69  & 142 & 68 \\
			$g$ & 270 & 131 & 73  & 132 & 70 \\
			$h$ & 270 & 120 & 63  & 121 & 66 \\
			\hline
			\hline
		\end{tabular*}
	\end{center}
\end{table}

Energy signals from the HPGe detectors, their associated bismuth-germanate Compton suppressors, and each individual photodiode were recorded and timestamped by an XIA Pixie-16 digital data acquisition system.
The data were acquired in list mode with no coincidence trigger condition, then processed to construct coincidence events using a coincidence window of $3$~$\mu$s.
The particle-$\gamma$ time-difference spectrum showed a clear prompt peak, so a timing cut was made around this peak with a width of $200$~ns, with random background subtractions from both sides.
Gamma-ray sources of \nuc{152}{Eu} and \nuc{207}{Bi} were used for energy and detection-efficiency calibrations.
Doppler-shift corrections to $\gamma$-ray energies were performed, assuming that the $\gamma$ rays were detected in coincidence with a beam particle.
The energy-calibrated, particle-gated $\gamma$-ray spectrum for each individual HPGe-photodiode pair was stored in a ROOT~\cite{Brun_1996} histogram for further analysis.
Due to their limited statistics, data from the different particle-gated $\gamma$-ray spectra were summed over the individual HPGe detectors to produce one spectrum for each photodiode.
Measured $\gamma$-ray yields were then corrected for HPGe detection efficiency and particle-$\gamma$ angular-correlation effects to produce an effective $4\pi$ $\gamma$-ray yield.

\section{RESULTS}
\begin{figure}[t]
	\includegraphics[width = 8.6 cm]{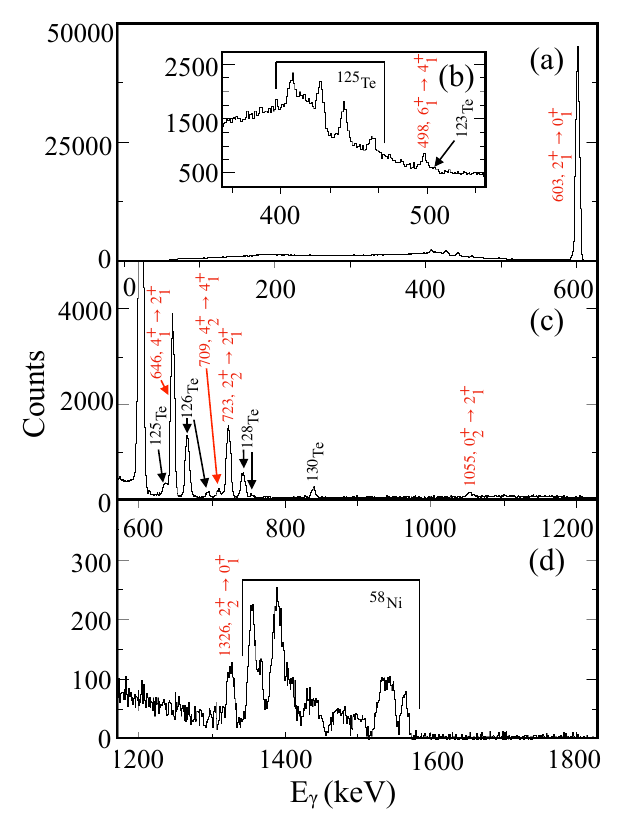}
	\caption{
		The particle-gated $\gamma$-ray spectrum collected during the measurement with a \nuc{58}{Ni} beam, summed over all detectors.
		Gamma rays detected following relaxation of Coulomb-excited \nuc{124}{Te} are labeled in red with energies and initial and final state spin-parities.
		Contaminant $\gamma$ rays from target or beam excitation are labeled in black with their origin.\label{fig:3}
	}
\end{figure}

\begin{figure}[h]
	\includegraphics[width = 8.6 cm]{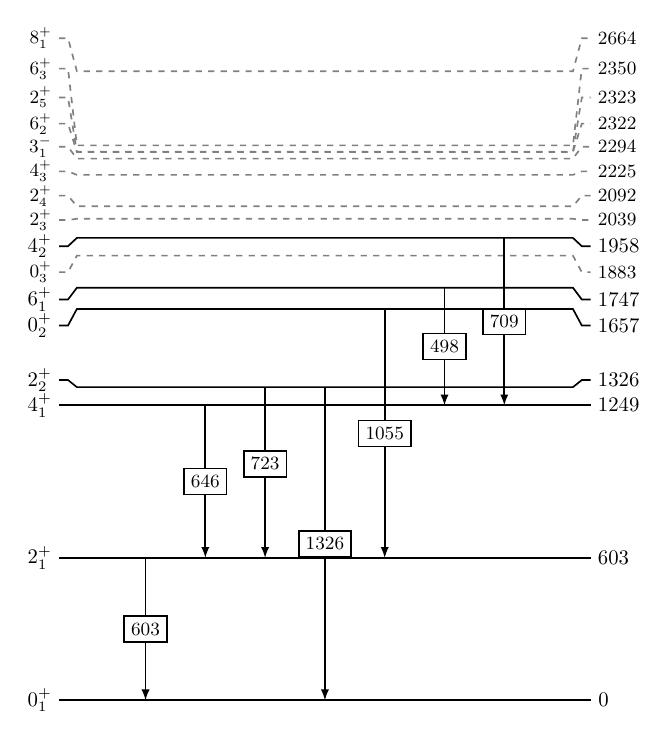}
	\caption{
		Low-lying partial level scheme of \nuc{124}{Te} showing the experimentally observed transitions along with levels included in the final analysis of the matrix elements. Level properties and transition energies (given in keV) are taken from Ref.~\cite{124Te}.\label{fig:4}
	}
\end{figure}

The total particle-gated, time-random-subtracted, $\gamma$-ray spectrum collected with the \nuc{58}{Ni} beam is shown in Fig.~\ref{fig:3}.
Eight transitions from the adopted level scheme~\cite{124Te} were identified; these are shown in the partial level scheme in Fig.~\ref{fig:4}.
The $4^+_2 \to 2^+_1$ transition at $1355$~keV was partially obscured by the Doppler-shifted $2^+_1 \to 0^+_1$, 1454.2-keV transition in \nuc{58}{Ni}, but its intensity can be inferred from the observed $4^+_2 \to 4^+_1$ transition and the known branching ratio~\cite{124Te,Hicks_2017}.
In the \nuc{16}{O}-beam data, only the 603-keV ($2^+_1 \to 0^+_1)$, 646-keV ($4^+_1 \to 2^+_1)$ and 723-keV ($2^+_2 \to 2^+_1)$ transitions were present. This is consistent with the observed excitation of the $2^+_1$, $2^+_2$, and $4^+_1$ states by Barrette \textit{et al.}~\cite{Barrette_1974} using a 42-MeV \nuc{16}{O} beam. The measured $\gamma$-ray yields, corrected for the relative HPGe detection efficiency and particle-$\gamma$ angular-correlation effects to produce an effective $4\pi$ array, are presented in Table~\ref{tab:yields}.

\begin{table}[t]
	\begin{center}
		\caption{
			Total number of counts measured for $\gamma$-ray transitions in \nuc{124}{Te} observed with each beam species, corrected for the relative HPGe detection efficiency and particle-$\gamma$ angular-correlation effects to produce the yield for an effective $4\pi$ array.
			Total yields normalized to the $2_1^+ \to 0_1^+$ transition in \nuc{124}{Te} are also included.\label{tab:yields}
		}
		\renewcommand{\arraystretch}{1.5}
		\begin{tabular*}{0.48\textwidth}{@{\extracolsep{\fill}}ccrr}
			\toprule
			 Beam & Transition & Counts & Yield\\
			\hline
			\nuc{16}{O} & $2^+_1 \rightarrow 0_1^+$  &  $1.23(2) \times 10^{7}$ & 1.0  \\
					   & $4^+_1 \rightarrow 2_1^+$  &  $1.73(14) \times 10^{5}$ & $1.42(11) \times 10^{-2}$  \\
					   & $2^+_2 \rightarrow 2_1^+$  &  $1.18(12) \times 10^{5}$ & $0.96(10) \times 10^{-2}$  \\
\hline
			\nuc{58}{Ni} & $2^+_1 \rightarrow 0_1^+$  &  $5.01(7) \times 10^{7}$ & 1.0  \\
					   & $4^+_1 \rightarrow 2_1^+$  &  $4.78(6) \times 10^{6}$ & $9.55(18) \times 10^{-2}$  \\
					   & $2^+_2 \rightarrow 2_1^+$  &  $2.34(4) \times 10^{6}$ & $4.68(11) \times 10^{-2}$  \\
					   & $2^+_2 \rightarrow 0_1^+$  &  $3.51(28) \times 10^{5}$ & $0.70(6) \times 10^{-2}$  \\
					   & $0^+_2 \rightarrow 2_1^+$  &  $2.73(65) \times 10^{5}$ & $0.55(13) \times 10^{-2}$  \\
					   & $6^+_1 \rightarrow 4_1^+$  &  $1.93(25) \times 10^{5}$ & $0.39(5) \times 10^{-2}$  \\
					   & $4^+_2 \rightarrow 4_1^+$  &  $2.01(30) \times 10^{5}$ & $0.40(6) \times 10^{-2}$  \\
			\hline
			\hline
		\end{tabular*}
	\end{center}
\end{table}

The semiclassical Coulomb-excitation code \textsc{Gosia}~\cite{Cline_2012} was used to extract reduced $E2$ matrix elements from the measured $\gamma$-ray yields.
Yields from both \nuc{58}{Ni} and \nuc{16}{O} beam data were included.
Stopping powers for the \textsc{Gosia} input were calculated using the program \textsc{SRIM}~\cite{ZIEGLER_2010}, and internal conversion coefficients for the relevant transitions were calculated using \textsc{BrIcc}~\cite{KIBEDI_2008}.

In \textsc{Gosia}, the matrix elements are fitted relative to a specific transition; in this case, the $2^+_1 \to 0^+_{1}$ transition in \nuc{124}{Te} was chosen.
Since the corresponding matrix element for that transition cannot be fitted simultaneously, a fixed value must be chosen for the minimization procedure.
A measured~\cite{Saxena_2014} value of $B(E2;0^+_1 \to 2^+_1) = 0.567(5)$~$e^2$b$^2$ was used to obtain the fixed matrix element, $\rme{2_1^+}{M(E2)}{0_1^+} = 0.753$~$e$b, where $M(E2)$ is the $E2$ operator and $\rme{I_f}{M(E2)}{I_i}$ denotes the reduced matrix element corresponding to the transition $I_i \to I_f$.
Independently adjusting this value to the limits that encompass the experimental uncertainty from Ref.~\cite{Saxena_2014} was found to change the fitted matrix elements by less than 0.5\%.
The $\rme{2_1^+}{M(E2)}{0_1^+}$ matrix element was also varied in the full correlated error calculation routine.

The set of measured yields was insufficient to uniquely determine the diagonal matrix elements, so these were also fixed in the fitting. The diagonal matrix elements are related to the spectroscopic quadrupole moments ($Q_s$)~\cite{Hamilton_1975} by:

\begin{equation}
	Q_s(I) = \rme{I}{M(E2)}{I} \sqrt{\frac{16\pi I (2I-1)}{5(2I+1)(I+1)(2I+3)}}~.
\end{equation}

The value of $\rme{2_1^+}{M(E2)}{2_1^+}$ was evaluated from the literature value of the spectroscopic quadrupole moment, $Q_s(2^+_1) = -0.45(5)$~$e$b~\cite{Stone_2021}. This value was adopted from Ref.~\cite{Bockisch_1976}, assuming a positive sign of the interference term given by:

\begin{equation}
	P_3 = \rme{0_1^+}{M(E2)}{2_1^+}\rme{2_1^+}{M(E2)}{2_2^+}\rme{2_2^+}{M(E2)}{0_1^+}~.
\end{equation}

It is not clear whether the sign of this interference term was determined in the original work of Bockisch~$et~al$.~\cite{Bockisch_1976}, and the present data do determine the $P_3$ sign. A negative interference term would give $Q_s(2^+_1)~=~-0.19$. The minimization of \textsc{Gosia} was performed using both values of $Q_s(2^+_1)$ and the variations in the results are incorporated into the quoted uncertainties. It is worth noting that the shell model predicts a negative interference term for \nuc{124}{Te} and a quadrupole moment, which is in better agreement with the associated value; however, further experimental studies are recommended. As no experimental values are known for $Q_s(2^+_2)$, $Q_s(4^+_1)$, $Q_s(4^+_2)$, and $Q_s(6^+_1)$, these were scaled from $Q_s(2^+_1)$ using the rotational approximation to set the intrinsic quadrupole moments ($Q_0$)~\cite{Bohr_1975}:

\begin{equation}
	Q_s = \frac{3K^2 - I(I+1)}{(I+1)(2I+3)}Q_0~.
\end{equation}

The \textsc{Gosia} analysis was also performed using values obtained from the shell model and the triaxial-rotor model~\cite{Davydov_1958}. However, the off-diagonal matrix elements of interest were found to be insensitive to the choice of diagonal matrix elements. As described in the following paragraphs, other unknown or uncertain matrix elements have a greater impact on the extracted results. 

Where they are available, spectroscopic data from other works are used to constrain the magnitudes of matrix elements for unobserved transitions. These include excited-state lifetimes, decay branching ratios and multipole mixing ratios summarized in Table~\ref{tab:spectroscopic-124te}. Matrix elements for transitions from the known $2^+_{2,3,4,5}$, $4^+_{2,3}$, and $3^-_1$ states were taken from $(n,n'\gamma)$ measurements~\cite{Hicks_2017}, with the exception of the adopted value for the $2^+_3 \to 2^+_1$ mixing ratio which is from Ref.~\cite{124Te}. These are summarized in Table~\ref{tab:fixedE2ME} and Table~\ref{tab:fixedme}. Inclusion of these data also constrained the matrix elements corresponding to some observed transitions. The fitted values of $\rme{2_2^+}{M(E2)}{2_1^+}$, $\rme{2_2^+}{M(E2)}{0_1^+}$, and $\rme{4_2^+}{M(E2)}{4_1^+}$ were influenced by the inclusion of excited-state lifetimes, while the fits to $\rme{6_1^+}{M(E2)}{4_1^+}$ and $\rme{0_2^+}{M(E2)}{2_1^+}$ were largely insensitive to the previous experimental data.

\begin{table}[h!]
	\begin{center}
		\caption{
			Summary of the excited-state lifetimes, transition branching ratios and $E2/M1$ mixing ratios used in the \textsc{Gosia} analysis. These values were taken from $(n,n'\gamma)$ measurements~\cite{Hicks_2017}, with the exception of the adopted value for the $2^+_3 \to 2^+_1$ mixing ratio from Ref.~\cite{124Te}. \label{tab:spectroscopic-124te}
		}
		\renewcommand{\arraystretch}{1.5}
		\begin{tabular*}{0.48\textwidth}{@{\extracolsep{\fill}}lr}
			\toprule
			 State & Lifetime (ps) \\
\hline
			$2^+_2$ & 0.85(17)   \\
			$4^+_2$ & 0.52(11)   \\
			$2^+_3$ & 0.88(11)   \\
			$2^+_4$ & 0.74(12)   \\
			$3^-_1$ & 0.15(2)   \\
			$2^+_5$ & 0.085(7)   \\
			$4^+_3$ & 0.34(1)   \\
\hline
			 Decay branch & Branching ratio ($\%$)\\
			\hline
			$2^+_2 \rightarrow 0_1^+$ / $2^+_2 \rightarrow 2_1^+$ &  0.149(12)  \\
			$4^+_2 \rightarrow 2_1^+$ / $4^+_2 \rightarrow 4_1^+$  & 0.902(26)  \\
			$4^+_2 \rightarrow 2_2^+$ / $4^+_2 \rightarrow 4_1^+$  & 0.0588(200)   \\
			$2^+_3 \rightarrow 0_1^+$ / $2^+_3 \rightarrow 2_1^+$  & 0.59(2)  \\
			$2^+_3 \rightarrow 4_1^+$ / $2^+_3 \rightarrow 2_1^+$  & 0.016(16)   \\
			$2^+_3 \rightarrow 2_2^+$ / $2^+_3 \rightarrow 2_1^+$  & 0.033(16)   \\
			$2^+_3 \rightarrow 0_2^+$ / $2^+_3 \rightarrow 2_1^+$  & 0.001(16)   \\
			$2^+_4 \rightarrow 0_1^+$ / $2^+_4 \rightarrow 2_1^+$  & 0.076(11)   \\
			$2^+_4 \rightarrow 2_2^+$ / $2^+_4 \rightarrow 2_1^+$  & 0.01(1)   \\
			$2^+_5 \rightarrow 0_1^+$ / $2^+_5 \rightarrow 2_1^+$  & 0.021(10)   \\
			$2^+_5 \rightarrow 2_2^+$ / $2^+_5 \rightarrow 2_1^+$  & 0.021(10)   \\
			$4^+_3 \rightarrow 2_1^+$ / $4^+_3 \rightarrow 4_1^+$  & 0.57(1)   \\
			$4^+_3 \rightarrow 2_2^+$ / $4^+_3 \rightarrow 4_1^+$  & 0.21(1)   \\
\hline
			 Transition & Mixing ratio \\
\hline
			$2^+_2 \rightarrow 2_1^+$  &  $-$3.4(3), $-$0.96(1)  \\
			$4^+_1 \rightarrow 4_1^+$  &   $-$0.36(3) \\
			$2^+_3 \rightarrow 2_1^+$  &    0.13(4)\footnotemark[1]\\
			$2^+_4 \rightarrow 2_1^+$  &    0.85(2)\\
			$2^+_4 \rightarrow 2_2^+$  &    0.58(92)\\
			$2^+_5 \rightarrow 2_1^+$  &    $-$0.03(9)\\
			$2^+_5 \rightarrow 2_2^+$  &    $-$0.36(22)\\
			$4^+_3 \rightarrow 4_1^+$  &    0.44(14)\\
			\hline
			\hline
		\end{tabular*}
	\end{center}
	\footnotetext[1]{Ref.~\cite{124Te}}
\end{table}

The adopted practice in Coulomb-excitation measurements is to include and investigate the effects of additional unobserved `buffer states' in the \textsc{Gosia} analysis. In cases where measured spectroscopic data are lacking, theoretical models may also be used if they are considered reliable. In the present study, results of General Collective Model calculations, described in detail in the discussion below, were used to estimate matrix elements for transitions from unobserved states ($8_1^+$, $6_{2,3}^+$) that could be connected to the $6_1^+$ level of interest. However, these calculated values are considered to be over-estimates of the true matrix-element magnitudes (see Section \ref{sec:discussion}). Their inclusion led to an over-prediction in the population of unobserved states in the calculations, thereby imposing limits on the extent by which the observed states can couple to them. A systematic evaluation of the sensitivity to these transitions was performed within the limitations imposed by the observed spectra. The measured matrix elements were largely insensitive to these unobserved buffer states, which were removed in the final \textsc{Gosia} analysis, and their effects were included as a source of uncertainty in the quoted values. \\

The measured matrix elements are also sensitive to the relative signs of the matrix elements adopted in the analysis. The phase of each excited-state wave function was chosen by fixing the sign for one matrix element of a transition from that state, and identifying the combination of matrix-element signs from other transitions that resulted in a global minimum for the analysis. Transitions within the ground-state band were assigned as `positive', as per convention, and transitions from other states were evaluated on a case-by-case basis. \\

\begin{table}[ht!]
	\begin{center}
		\caption{
			List of $E2$ matrix elements that were fixed during the \textsc{Gosia} analysis. Unless otherwise stated, the values are derived from the adopted lifetimes, energies, branching ratios, and $E2/M1$ mixing ratios from $(n,n'\gamma)$ measurements~\cite{Hicks_2017}. Matrix elements with signs fixed as `positive' are marked with~$\oplus$.
			}  
			\label{tab:fixedE2ME}
		\renewcommand{\arraystretch}{1.5}
		\begin{tabular*}{0.48\textwidth}{@{\extracolsep{\fill}}cr}
			\toprule
			\hline
			 \multicolumn{2}{c}{$E2$ matrix elements (eb)} \\
\hline
		$\rme{2_1^+}{M(E2)}{0_1^+}$ & $\oplus$0.753 \\
		$\rme{2_3^+}{M(E2)}{0_1^+}$ & $\pm$0.069 \\
		$\rme{2_4^+}{M(E2)}{0_1^+}$ & $\pm$0.031 \\
		$\rme{2_5^+}{M(E2)}{0_1^+}$ & $\pm$0.038 \\
		$\rme{2_1^+}{M(E2)}{2_1^+}$ & $-$0.594\footnotemark[1] \\
		$\rme{2_3^+}{M(E2)}{2_1^+}$ & $\oplus$0.028 \\
		$\rme{2_4^+}{M(E2)}{2_1^+}$ & $\oplus$0.171 \\
		$\rme{2_5^+}{M(E2)}{2_1^+}$ & $\pm$0.017 \\
		$\rme{4_3^+}{M(E2)}{2_1^+}$ & $\pm$0.144 \\
		$\rme{4_1^+}{M(E2)}{4_1^+}$ & $-$0.760\footnotemark[2] \\
		$\rme{2_3^+}{M(E2)}{4_1^+}$ & $\pm$0.122 \\
		$\rme{4_3^+}{M(E2)}{4_1^+}$ & $\oplus$0.273 \\
		$\rme{2_2^+}{M(E2)}{2_2^+}$ & 0.594\footnotemark[2] \\
		$\rme{2_3^+}{M(E2)}{2_2^+}$ & $\pm$0.224 \\
		$\rme{2_4^+}{M(E2)}{2_2^+}$ & $\pm$0.073 \\
		$\rme{2_5^+}{M(E2)}{2_2^+}$ & $\oplus$0.106 \\
		$\rme{4_3^+}{M(E2)}{2_2^+}$ & $\pm$0.665 \\
		$\rme{2_3^+}{M(E2)}{0_2^+}$ & $\pm$0.736 \\
		$\rme{6_1^+}{M(E2)}{6_1^+}$ & $-$0.904\footnotemark[2] \\
\hline
\hline
		\end{tabular*}
	\end{center}
	\footnotetext[1]{Ref.~\cite{Stone_2021}.}
	\footnotetext[2]{Rotational approximation based on $Q_s(2^+_1)$.}
\end{table}

\begin{table}[ht!]
	\begin{center}
		\caption{
			List of $E1,~E3,$~and$~M1$ matrix elements that were fixed during the \textsc{Gosia} analysis. Unless otherwise stated, the values are derived from the adopted lifetimes, energies, branching ratios, and $E2/M1$ mixing ratios from $(n,n'\gamma)$ measurements~\cite{Hicks_2017}, with the exception of the adopted value for the $2^+_3 \to 2^+_1$ mixing ratio from Ref.~\cite{124Te}. \label{tab:fixedme}
		}
		\renewcommand{\arraystretch}{1.5}
		\begin{tabular*}{0.48\textwidth}{@{\extracolsep{\fill}}cr}
			\toprule
			 \multicolumn{2}{c}{$E1$ matrix elements (eb$^{1/2}$)} \\

\hline
		$\rme{3_1^-}{M(E1)}{2_1^+}$ & 0.00741  \\
		$\rme{3_1^-}{M(E1)}{4_1^+}$ & 0.00351 \\
		$\rme{3_1^-}{M(E1)}{2_2^+}$ & 0.00318 \\
			\hline
			 \multicolumn{2}{c}{$E3$ matrix elements (eb$^{3/2}$)} \\
\hline
		$\rme{3_1^-}{M(E3)}{0_1^+}$ & 0.414 \\
			\hline
			 \multicolumn{2}{c}{$M1$ matrix elements ($\mu_N$)} \\
\hline
		$\rme{2_3^+}{M(M1)}{2_1^+}$ & 0.258 \\
		$\rme{2_4^+}{M(M1)}{2_1^+}$ & 0.249 \\
		$\rme{2_5^+}{M(M1)}{2_1^+}$ & $\mp$0.792 \\
		$\rme{4_3^+}{M(M1)}{4_1^+}$ & 0.871 \\
		$\rme{2_4^+}{M(M1)}{2_2^+}$ & $\pm$0.080 \\
		$\rme{2_5^+}{M(M1)}{2_2^+}$ & $-0.245$ \\
\hline
\hline
		\end{tabular*}
	\end{center}
\end{table}

The set of matrix elements measured in this work is summarized in Table~\ref{tab:measuredme}. The extracted value of $\rme{2_2^+}{M(E2)}{0_1^+}$ was found to be highly sensitive to the $E2/M1$ mixing ratio for the $2^+_2 \to 2^+_1$ transition. Two possible values of this mixing ratio ($-0.96(1)$ and $-0.49^{+5}_{-3}$) are reported from the $(n,n'\gamma)$ measurements~\cite{Hicks_2017}, both of which disagree with the earlier adopted value ($-3.4(3)$) derived from $\gamma-\gamma$ angular correlations in $\beta$-decay experiments~\cite{124Te}. Using the mixing ratio from Ref.~\cite{124Te}, it was not possible to achieve a satisfactory fit to the measured yields while also reproducing the lifetime of the $2^+_2$ state that was reported in Ref.~\cite{Hicks_2017}. Alternatively, using the two mixing ratios from Ref.~\cite{Hicks_2017} resulted in two different values for each of $\rme{2_2^+}{M(E2)}{2_1^+}$ and $\rme{2_2^+}{M(E2)}{0_1^+}$. To address this discrepancy, we have re-examined existing data from measurements made by our group for the PhD thesis of B.~J.~Coombes~\cite{Coombes_thesis}, results of which will be published separately. These measurements used similar photodiode particle detectors and HPGe $\gamma$-ray detectors on a neighboring beamline to collect particle-$\gamma$ angular correlation data from \nuc{58}{Ni} on \nuc{124}{Te} at energies near, but slightly above, the safe Coulomb-excitation criterion. These data rule out the case of $\delta_{2^+_2 \to 2^+_1} = -0.49$, although they cannot distinguish between $\delta_{2^+_2 \to 2^+_1} = -0.96$ and $\delta_{2^+_2 \to 2^+_1} = -3.4$. Since this discrepancy remains unresolved, the fitted matrix elements obtained using each of the two possible mixing ratios are given below. All other measured matrix elements were found to be independent of $\delta_{2^+_2 \to 2^+_1}$. The $\rme{4_1^+}{M(E2)}{2_1^+}$ value was found to be sensitive to the sign combination of the $4^+_2 \to 2^+_1$ and $4^+_2 \to 2^+_2$ transitions. The value determined with $\rme{4_2^+}{M(E2)}{2_1^+}<0$ and $\rme{4_2^+}{M(E2)}{4_1^+}>0$ is consistent with the results of Ref.~\cite{Saxena_2014}, whereas the other result is 2$\sigma$ away; we have included both values in the following discussion.\\


The off-diagonal reduced matrix elements of the $E2$ operator are related to the reduced transition strengths by: \\[-1cm]

\begin{align}
	\rme{I_f}{M(E2)}{I_i}^2 = (2 I_i + 1 ) B(E2; I_i \to I_f).
\end{align}

The measured matrix elements and corresponding $B(E2)$ transition strengths are given in Table~\ref{tab:measuredme}.
Uncertainties provided include those from the \textsc{Gosia} correlated error routine, and account for the possible variation of all matrix elements, including those which were fixed in the fitting, such as the quadrupole moments as well as variation from the relative signs of matrix elements, inclusion of buffer states, and unknown diagonal matrix elements.
Previous measurements are included where they exist; this is the first measurement of the $B(E2;6^+_1 \to 4^+_1)$ value.

\begin{table*}[t]
	\begin{ruledtabular}
		\caption{
			Summary of the matrix elements, $\rme{I_f}{M(E2)}{I_i}$, of transitions that connect levels $I_i$ and $I_f$ in \nuc{124}{Te} measured in this work.
			The $\gamma$-ray energies ($E_{\gamma}$) are from Ref.~\cite{124Te}, and multipole mixing ratios ($\delta _ {2^+_2 \to 2^+_1}$) are discussed in the text.
			Matrix elements fitted with $\delta_{2^+_2 \to 2^+_1} = -3.4(3)$~\cite{124Te} do not reproduce the $2^+_2$-state lifetime from $(n,n'\gamma)$~\cite{Hicks_2017}.
			The transition strengths, $B(E2; I_i \to I_f)$, corresponding to the matrix elements are given in Weisskopf units (W.u.), along with previous measurements where they exist. Matrix elements with signs fixed as `positive' are marked with~$\oplus$.
			\label{tab:measuredme}
		}
		\renewcommand{\arraystretch}{1.5}
		\begin{tabular}{ccccccccc} 
			$I_i$   	&	  $I_f$    & $E_{\gamma}$ 	& $\delta_{2^+_2 \to 2^+_1}$ & $\rme{I_f}{M(E2)}{I_i}$ & \multicolumn{4}{c}{$B(E2; I_i \to I_f)$~(W.u.)}\\ \cline{6-9}
			& 	  &	(keV) &  &($e$b)   & This work	& Ref.~\cite{Saxena_2014}  & Ref.~\cite{Hicks_2017} & Ref.~\cite{124Te}	  \\[0.05cm]
			\hline
			$2^+_1$ & $0^+_1$ & 603  &                                     &                          & [30.9(3)] \footnotemark[1]   & 30.9(3)      &                     & 31.1(5)  \\
			$4^+_1$ & $2^+_1$ & 646  &                                     & $\oplus$~1.04(7) \footnotemark[3] & 33(4)            & 35.9(17) &                     & $97^{+54}_{-49}$ \footnotemark[2]       \\
			        &         &      &                                     & $\oplus$~1.23(7) \footnotemark[4] & 46(5)            &          &                     &                                         \\
			$2^+_2$ & $2^+_1$ & 723  & $-3.4(3)$~\cite{124Te}              & $\oplus$~$1.04^{+6}_{-7}$        & 53(7)            & 34(5)    &                     &       \\
			          &         &      & $-0.96(1)$~\cite{Hicks_2017}        & $\oplus$~0.88(6)                 & 42(6)            &          & $55.5^{+109}_{-99}$ &       \\
			$2^+_2$ & $0^+_1$ & 1326 & $-3.4(3)$~\cite{124Te}              & $\pm0.086^{+9}_{-10}$              & 0.38(5)          & 0.22(6)  &                     & $0.49^{+5}_{-10}$ \\
			          &         &      & $-0.96(1)$~\cite{Hicks_2017}        & $\pm0.095^{+9}_{-10}$      & 0.6(1)           &          & $0.83^{+23}_{-16}$  &       \\
			$6^+_1$ & $4^+_1$ & 498  &                                     & $\oplus$~$1.14^{+17}_{-19}$       & 27(9)            &          &                     &       \\
			$0^+_2$ & $2^+_1$ & 1055 &                                     & $\oplus$~$0.20^{+5}_{-7}$      & 11(6)            &          & $14.3(29)$          & 20(4) \\
			$4^+_2$ & $4^+_1$ & 709  &                                     & $\oplus$~$0.85^{+38}_{-32}$       & $22^{+24}_{-13}$         &          & $14.1^{+30}_{-28}$  &
		\end{tabular}
	\end{ruledtabular}
 			\footnotetext[1]{Adopted normalization value.}
			\footnotetext[2]{Calculated from the adopted lifetime. The value given in Ref.~\cite{124Te} is $B(E2) = 97.529(4)$ W.u., with the reported uncertainty evidently being incorrect.}
            \footnotetext[3]{Determined with $\rme{4_2^+}{M(E2)}{2_1^+}>0$ and $\rme{4_2^+}{M(E2)}{2_2^+}<0$.}
            \footnotetext[4]{Determined with $\rme{4_2^+}{M(E2)}{2_1^+}<0$ and $\rme{4_2^+}{M(E2)}{2_2^+}>0$.}
\end{table*}

\section{DISCUSSION}\label{sec:discussion}

\subsection{Comparison with previous work}
As discussed above, previous measurements~\cite{124Te} of transition strengths in \nuc{124}{Te} are sparse and there is some inconsistency.
Previous work of relevance to the present study includes Coulomb excitation by particle-spectroscopy~\cite{Barrette_1974} and lifetime measurements~\cite{Borner_1993, Doll_2000}.
However, the work of Saxena \textit{et al.}~\cite{Saxena_2014} is the most relevant as it is also a Coulomb-excitation study with \nuc{58}{Ni} beams at a similar bombarding energy (175 MeV) impinging on thin Te-enriched targets on a carbon backing.

The $B(E2)$ values measured in the present work are consistent with those of Saxena \textit{et al.}~\cite{Saxena_2014} for the $4^+_1 \to 2^+_1$ transition when the $\rme{4_2^+}{M(E2)}{4_1^+}<0$ sign is adopted. The value obtained using the other sign convention is within 2$\sigma$ of this result, and it is consistent with the $B(E2)$ derived from the adopted lifetime~\cite{124Te} (noting the large uncertainty). A discrepancy exists between the present $2^+_2 \to 2^+_1$ transition strength and previous measurements. This matrix element is highly sensitive to the assumed relative phases of matrix elements in the analysis. 

The $B(E2; 2^+_2 \to 0^+_1)$ values are also consistent when $\delta_{2^+_2 \to 2^+_1}=-3.4(3)$~\cite{124Te} is used.
However, in the Coulomb-excitation analysis, this transition's matrix-element value is strongly dependent on the multipole mixing ratio of the $2^+_2 \to 2^+_1$ transition.
A new independent measurement of this mixing ratio is warranted to resolve the discrepant reported values~\cite{124Te,Hicks_2017}.
An independent measurement of the $2^+_2$ lifetime by the recoil-distance method, or by Doppler shift with high-velocity recoils, is also recommended as the Coulomb-excitation data together with the mixing ratio from $\beta$ decay suggest a longer lifetime.

\subsection{$\boldsymbol{B(E2)}$ ratios and model benchmarks}
To address open questions concerning the nature of the low-lying structures observed in midshell tellurium isotopes, we turn our focus to the yrast states in \nuc{124}{Te} and begin with comparisons to various model benchmarks concerning $B(E2)$ ratios.
Although these benchmarks are oversimplifications, they are frequently the starting point for the characterization of low-energy nuclear structure.

\begin{figure}[t!]
	\includegraphics[width = 8.6 cm]{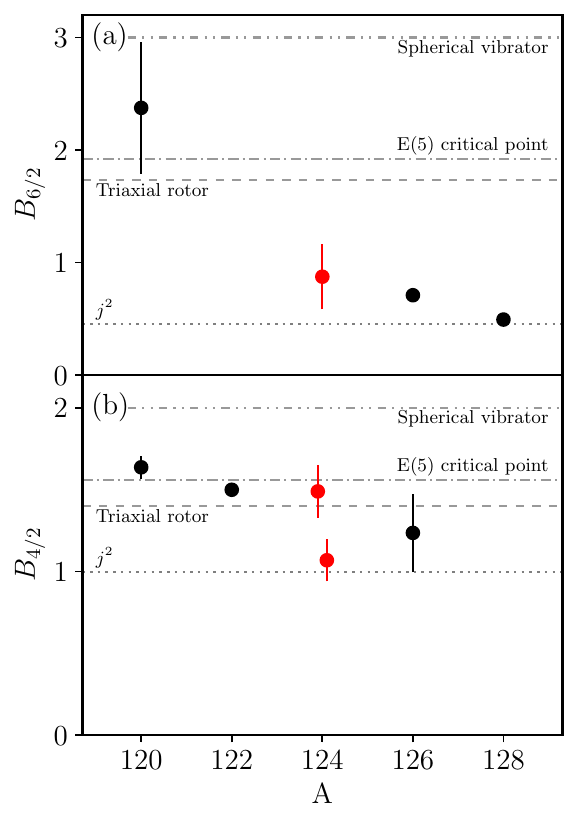}
	\caption{
		Ratios of the (a) $B(E2; 6^+_1 \to 4^+_1)$ and (b) $B(E2; 4^+_1 \to 2^+_1)$ values to the corresponding $B(E2; 2^+_1 \to 0^+_1)$ value, denoted $B_{6/2}$ and $B_{4/2}$, respectively, in even-even \nuc{120--128}{Te}.
		Experimental values for \nuc{124}{Te} (red circles) use $B(E2; 6^+_1 \to 4^+_1)$ and $B(E2; 4^+_1 \to 2^+_1)$ from this work and $B(E2;2^+_1 \to 0^+_1)$ from Ref.~\cite{Saxena_2014} as included in the \textsc{Gosia} fit.
		Experimental values for other isotopes (black circles) are from Refs.~\cite{Saxena_2014,Hicks_2005,Stokstad_1967,126Te,128Te}.
		Predictions from the spherical-vibrator model~\cite{Bohr_1953} (dash-dot-dot line), E(5) critical-point symmetry~\cite{Clark_2004} (dash-dot line), Davydov-Filippov model~\cite{Davydov_1958} (dashed line) and $j^2$ model (dotted line) are also shown.\label{fig:5}
	}
\end{figure}

Figure~\ref{fig:5} presents $B(E2; 6^+_1 \to 4^+_1)$ and $B(E2; 4^+_1 \to 2^+_1)$ values as a ratio to the corresponding $B(E2;2^+_1 \to 0^+_1)$ value for even-even \nuc{120--128}{Te}.
These ratios are hereafter referred to as $B_{6/2}$ and $B_{4/2}$, respectively.
The measured $B(E2)$ values are from the present and previous work, including data for neighboring isotopes, where available~\cite{Saxena_2014,Hicks_2005,Stokstad_1967,126Te,128Te}.
The $B(E2)$ ratios are compared to a range of collective-model limits and, from the single-particle perspective, the $j^2$ seniority limit, which considers the yrast states up to $I=6$ as rising from the proton $g_{7/2}^2$ configuration.
(See, for example, Ref.~\cite{Heyde_1990} for the relevant formulae.)

Three collective-model cases are shown: the spherical vibrational model, the triaxial-rotor model, and the E(5) critical-point symmetry of the Bohr collective model.

The vibrational model for spherical nuclei~\cite{Bohr_1953} predicts low-energy states arising from harmonic oscillations about a spherical ground-state shape, corresponding to the quadrupole phonon states shown on the left side of Fig.~\ref{fig:1}. The transition strengths from the highest-spin state in each phonon multiplet can be simply expressed relative to the transition from the first phonon:
\vspace{-0.3cm}
\begin{equation}
	B(E2;I=2n \to I = 2(n-1)) = nB(E2;2^+_1 \to 0^+_1),
\end{equation}

\noindent
where $n$ is the phonon number.
The data in Fig.~\ref{fig:5} show no agreement with the spherical vibrational model, despite the common description of these nuclei as vibrational, including the very recent work of Ref.~\cite{Gupta_2023}.
In \nuc{120}{Te}, $B_{6/2}$ may be consistent with the vibrational limit, but the uncertainty is large, and the $B_{4/2}$ value falls short of the vibrator value.

Saxena \textit{et al.}~\cite{Saxena_2014} point to rotational behavior in \nuc{120,122,124}{Te} based on their measured $B_{4/2}$ ratios.
They conclude that these nuclei are soft triaxial rotors.
As a benchmark, we use the Davydov-Filippov model~\cite{Davydov_1958}, which describes a nucleus with rigid triaxial deformation.
The $B(E2)$ ratios in this model depend on the triaxiality parameter ($\gamma$) but not the deformation ($\beta$).
The $B(E2)$ ratios for this model shown in Fig.~\ref{fig:5} were calculated using the code from Ref.~\cite{DFcode} with $\gamma=25^{\circ}$ and $\beta=0.17$, which reproduces the empirical relationship between $Q_s(2^+_1)$ and $B(E2;2^+_1 \to 0^+_1)$ in \nuc{124}{Te}.
At $\gamma = 25^{\circ}$, $B_{4/2}=1.426$ and $B_{6/2} = 1.781$, while for the symmetric rotor $\gamma = 0^{\circ}$, $B_{4/2}=30/21 \simeq 1.429$ and $B_{6/2} = 225/143 \simeq 1.574$.
Thus, the triaxiality has little impact on $B_{4/2}$ but increases $B_{6/2}$ by about 13\%.

The E(5) critical-point symmetry marks the transition between the spherical-vibrator limit and the triaxial-rotor limit of the Bohr collective model, based on a simplification of the potential $V(\beta,\gamma)$.
The $B(E2)$ ratios for the E(5) critical point symmetry with a linear quadrupole operator are $B_{4/2}=1.68$ and $B_{6/2} = 2.21$~\cite{Iachello_2000}.
Clark \textit{et al.} give somewhat smaller ratios corresponding to the use of a quadratic quadrupole operator, namely $B_{4/2}=1.56$ and $B_{6/2} = 1.92$~\cite{Clark_2004}.
The latter values from Clark \textit{et al.}, which are about 10\% higher than the Davydov-Filippov model, are shown in Fig.~\ref{fig:5}.

In Fig.~\ref{fig:5}, it is seen that the $B_{4/2}$ ratios of both the Davydov-Filippov and the E(5) models are in reasonable agreement with the data for \nuc{120,122,124}{Te}, whereas $B_{6/2}$ agrees for \nuc{120}{Te} (albeit with a large uncertainty) but not for \nuc{124}{Te}.
The new data on $B_{6/2}$ indicate that \nuc{124}{Te} is not a good candidate for the E(5) model.

\subsection{General collective model}
The general collective model (GCM) is a phenomenological model of low-excitation collective nuclear states \cite{Hess1980,POHess1981,Eisenberg-Greiner,Troltenier1991}.
It makes no distinction between protons and neutrons and treats the nuclear states as the excitations (vibrations and rotations) of an incompressible liquid drop.
The version of the model used here \cite{Troltenier1991} has two parameters to define the kinetic energy and six parameters to define the potential $V(\beta, \gamma)$ as an expansion in $\beta$ up to sixth order (i.e.\ $\beta^6$).
These eight parameters are determined by fitting the low-excitation spectrum and the $E2$ properties of the lowest few states (i.e.\ $E2$ transition strengths and quadrupole moments).

\begin{figure*}[th!]
	\begin{center}
		\includegraphics[width = 17.8 cm]{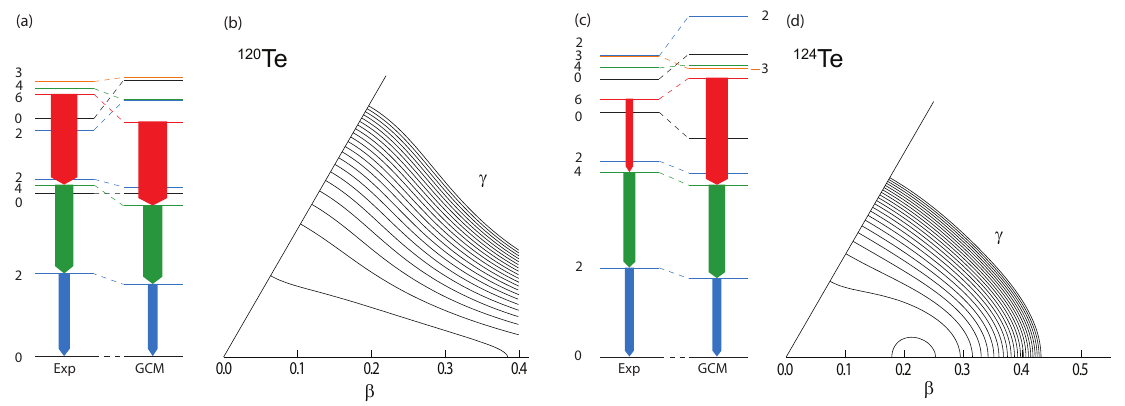}
		\caption{
			General collective model calculations for \nuc{120}{Te} and \nuc{124}{Te} comparing the experimental and theoretical levels in panels (a) and (c), respectively, and showing the potential energy surfaces in panels (b) and (d), respectively.
            Potential energy surface parameters are given in Table~\ref{tab:gcmparam}.
			The arrows indicate the $B(E2)$$\downarrow$ values for the 2$^+_1$, 4$^+_1$ and 6$^+_1$ states. For \nuc{124}{Te}, an average of the two $B(E2; 4^+_1 \to 2^+_1)$ values presented in Table~\ref{tab:measuredme} and discussed in the text is shown here for illustration.
			A detailed comparison of experimental and theoretical $E2$ transition strengths is given in Table~\ref{tab:gcmBE2}.
   		  Note that the GCM significantly overestimates $B(E2; 6^+_1 \to 4^+_1)$ in \nuc{124}{Te}.
            \label{fig:6}
		}
	\end{center}
\end{figure*}

In the calculations presented here, the lowest states up to $3^+_1$ and $6^+_1$ (10 states including the ground state) are included in the fit, along with the $B(E2)$ values among these low-excitation states and the quadrupole moment of the first excited state $Q_s(2^+_1)$, where available. The $P3$ kinetic-energy parameter was set to zero, leaving seven free parameters. 

It is noteworthy that there is no separate parameter to set the scale of the $B(E2)$ values.
However, the model is phenomenological with a considerable number of parameters (i.e.\ 8).
It is used here to assess whether the low-lying states of \nuc{120}{Te} and \nuc{124}{Te} can be described as purely collective structures, or not.
The parametrization of the potential is general enough to describe vibrational, rotational, triaxial, and soft nuclei, as well as shape coexistence.
Some caution is necessary, as sometimes rather different potential energy surfaces can give comparable fits to the available data.
Thus, a successful fit does not imply a unique interpretation of the structure of the nucleus.
However, the model's failure to fit data, especially $E2$ transition strengths, can strongly indicate that the excitations are not truly collective.

\begin{table}[t]
	\begin{center}
		\caption{
			Parameters employed in the GCM calculations \cite{Troltenier1991} .\label{tab:gcmparam}
		}
		\begin{tabular*}{0.48\textwidth}{@{\extracolsep{\fill}}ccccccccc}
			\toprule\\[-0.32cm]
			Isotope & $B2$ & $P3$ & $C2$ & $C3$ & $C4$ & $C5$ & $C6$  & $D6$  \\
			\hline \\[-0.3cm]
			\nuc{120}{Te} & 59.4 &  0 & $-18.2$ & 110 & 1245 & 3385 & $-25190$ & 17140 \\
			\nuc{124}{Te} & 58.2 &  0 & $-35.7$ & 208 & 1466 & 3254 & $-9132$  & 26550 \\
			\hline
			\hline
		\end{tabular*}
	\end{center}
\end{table}

\begin{table}[t]
	\begin{center}
		\caption{
			Comparison of general collective model (GCM) and experimental $B(E2)$ values in \nuc{120}{Te} and \nuc{124}{Te}.
			Experimental data are from Refs.~\cite{120Te,124Te,Saxena_2014} and the present work.\label{tab:gcmBE2}
		}
		\begin{tabular*}{0.48\textwidth}{@{\extracolsep{\fill}}cccc}
			\toprule\\[-0.32cm]
			Isotope & Transition & \multicolumn{2}{c}{$B(E2)$ W.u.}\\
			\cline{3-4} \\[-0.3cm]
			& & Exp & GCM  \\
			\hline\\[-0.3cm]
			\nuc{120}{Te} & $ 2^+_1 \to 0^+_1$    & 37.9(11)   & 32.7 \\
			& $ 4^+_1 \to 2^+_1$    & 62(2)      & 65.7 \\
			& $ 6^+_1 \to 4^+_1$    & 90(22)     & 97.9 \\
			& $ 2^+_2 \to 2^+_1$    & 46.1(23)   & 39.2 \\
			& $ 2^+_2 \to 0^+_1$    & 0.56(3)    & 0.05 \\
			\\
			\nuc{124}{Te} & $ 2^+_1 \to 0^+_1$    & 30.9(3)    & 30.7 \\
			& $ 4^+_1 \to 2^+_1$    & 33(4)\footnotemark[1], 46(5)\footnotemark[2] & 55.7 \\
			& $ 6^+_1 \to 4^+_1$    & 27(9)      & 74.9 \\
			& $ 2^+_2 \to 2^+_1$    & 53(7)\footnotemark[3], 42(6)\footnotemark[4]       & 37.0 \\
			& $ 2^+_2 \to 0^+_1$    & 0.38(5)\footnotemark[3], 0.6(1)\footnotemark[4]     & $9 \times 10^{-4}$ \\
			& $ 0^+_2 \to 2^+_1$    & 11(6)      & 41.9 \\
			& $ 4^+_2 \to 4^+_1$    & 14(3)      & 23.0\\
			\hline
			\hline
			\vspace{-0.8cm}
		\footnotetext[1]{The value corresponding to the minimization with $\rme{4_2^+}{M(E2)}{4_1^+}<0$ is shown here.}
		\footnotetext[2]{The value corresponding to the minimization with $\rme{4_2^+}{M(E2)}{4_1^+}>0$ is shown here.}
		\footnotetext[3]{The values corresponding to the adopted mixing ratio in Ref.~\cite{124Te} are shown here.}
		\footnotetext[4]{The values corresponding to the adopted mixing ratio in Ref.~\cite{Hicks_2017} are shown here.}

		\end{tabular*}
	\end{center}
\end{table}

The GCM parameter sets for \nuc{120}{Te} and \nuc{124}{Te} are given in Table~\ref{tab:gcmparam}; Table~\ref{tab:gcmBE2} compares the experimental and theoretical $B(E2)$ values.
The corresponding level schemes and potential energy surfaces are shown in Fig.~\ref{fig:6}. 
In addition to the $E2$ transitions, in \nuc{124}{Te}, the GCM gives $Q_s(2^+_1) = -0.45$~$e$b, in agreement with experiment, namely $Q_s(2^+_1) = -0.45(5)$~$e$b.
The \nuc{120}{Te} parameter set gives $Q_s(2^+_1) = -0.55$~$e$b, in reasonable agreement with the trend in $Q_s(2^+_1)$ values, although no experimental value is available.
Overall, the description of \nuc{120}{Te} by the GCM is satisfactory, although it can be noted that the absolute value of the $B(E2; 2^+_1 \to 0^+_1)$ is lower than experiment by 5 standard deviations.
The reason is that the model attempts to make the $B(E2)$ ratios $B_{4/2}$ and $B_{6/2}$ compatible with the vibrational model as suggested by the level sequence.

In the case of \nuc{124}{Te}, however, the GCM fit has the correct magnitude for the $B(E2; 2^+_1 \to 0^+_1)$, but it does not describe the $B(E2)$ strengths observed for the decays of the $4^+_1$ and $6^+_1$ states, which fall short by $(41 \pm7)\%$ or $(17 \pm 9)\%$ and $(64 \pm 12)\%$ of the GCM values, respectively.
The results imply that these states are not simple collective excitations, and moving from $N = 82$ toward the midshell, collectivity is still emerging at \nuc{124}{Te}.
This overestimate of the $E2$ decay strength is particularly the case for the $4^+_1$ state, and more so for the $6^+_1$ state.
These observations are consistent with the view in Ref.~\cite{Lee_1991} that the $6^+_1$ state has a strong two-proton (seniority) structure while the $4^+_1$ state is a mix of collective and seniority structures.
Shell-model calculations in the next section explore this possibility.

Finally, to conclude this section, some remarks are made on the transitions from the non-yrast states.
Without a well-constrained $2^+_2 \to 2^+_1$ mixing ratio in \nuc{124}{Te}, limited information can be extracted from comparisons between new and existing $B(E2; 2^+_2 \to 0^+_1)$ values.
Nonetheless, for each reported mixing ratio, we obtain a value for $B(E2;2^+_2 \to 0^+_1)$ that is small but nonzero (see Table~\ref{tab:measuredme}).
This is contrary to the predictions of the vibrational model, in which this transition is forbidden.
Note that the GCM calculations also underestimate this transition strength in both \nuc{120}{Te} and \nuc{124}{Te}.

In addition, our measurement of $B(E2;0^+_2 \to 2^+_1)$ is substantially below the prediction given by the vibrational model, for which it should be twice the value of $B(E2;2^+_1 \to 0^+_1)$.
The GCM is comparatively in better agreement with experiment (a ratio to $B(E2;2^+_1 \to 0^+_1)$ of 1.4 rather than 2), but still it greatly overestimates the $E2$ decay strength from the $0^+_2$ state, which experimentally is less than half of the $B(E2;2^+_1 \to 0^+_1)$ value.
This small $B(E2)$ value may be a signal of the shape coexistence proposed on the basis of the level systematics~\cite{RIKOVSKA_1989}, but more data are required to draw a firm conclusion.

\subsection{Shell-model calculations}
The $j^2$-model $B(E2)$ ratios shown in Fig.~\ref{fig:5}, which describe the $0^+_1$, $2^+_1$, $4^+_1$, and $6^+_1$ states as arising from alternative couplings of the two valence protons in the $0g_{7/2}$ orbit, can provide a realistic approximation only at $N=82$, i.e.\ for \nuc{134}{Te}.
In fact, the measured $E2$ strengths in \nuc{134}{Te} follow this simple formulation rather well~\cite{Stuchbery_2013}.
Away from $N=82$, however, shell-model calculations in a more extended basis space are required.

Large-scale shell-model calculations were performed with the code \textsc{KSHELL}~\cite{Shimizu_2019} using a \nuc{100}{Sn} core.
Both protons and neutrons occupied the orbits $0g_{7/2}$, $1d_{5/2}$, $1d_{3/2}$, $2s_{1/2}$, and $0h_{11/2}$.
The effective shell-model Hamiltonian, including the single-particle energies and two-body matrix elements, was derived from the CD-Bonn potential as described by Coraggio \textit{et al.}~\cite{Coraggio_2017}.
As in previous work~\cite{Gray_2020}, effective charges of $e_p = 1.7 e$ and $e_n = 0.9 e$ were determined empirically from the semimagic nuclei \nuc{134}{Te} and \nuc{128}{Sn}, respectively.

\begin{figure*}[ht]
	\includegraphics[width=17.5cm]{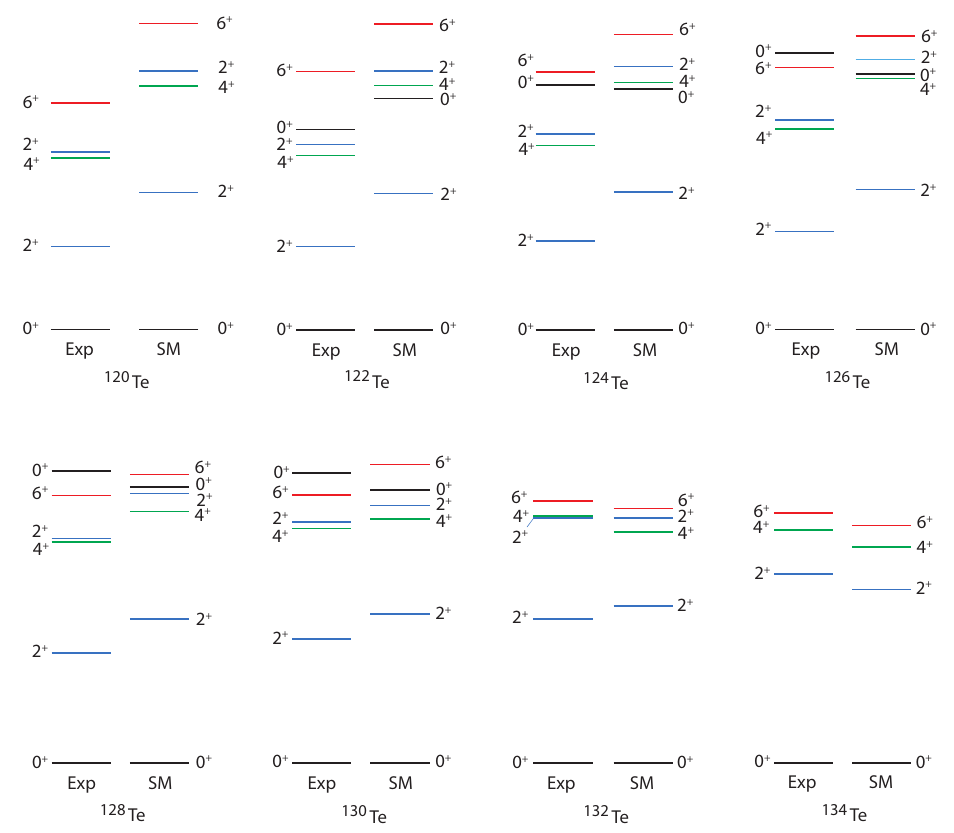}
	\caption{
		Comparison of shell-model calculations and experimental energies~\cite{120Te,122Te,124Te,126Te,128Te,130Te,132Te,134Te} for low-excitation-energy states (up to the first $6^+$ state) in the even-even isotopes from \nuc{120}{Te} to \nuc{134}{Te}.\label{fig:7}
	}
\end{figure*}

Figure~\ref{fig:7} compares the experimental and shell-model excitation energies for the low-excitation states in \nuc{120--134}{Te} up to the $6^+_1$ state.
The agreement between theory and experiment is satisfactory, although there is a clear trend for the calculated excitation energies to increasingly exceed experiment as the number of valence neutron holes increases toward \nuc{120}{Te}.

\begin{figure}[ht]
	\includegraphics[width = 8.6cm]{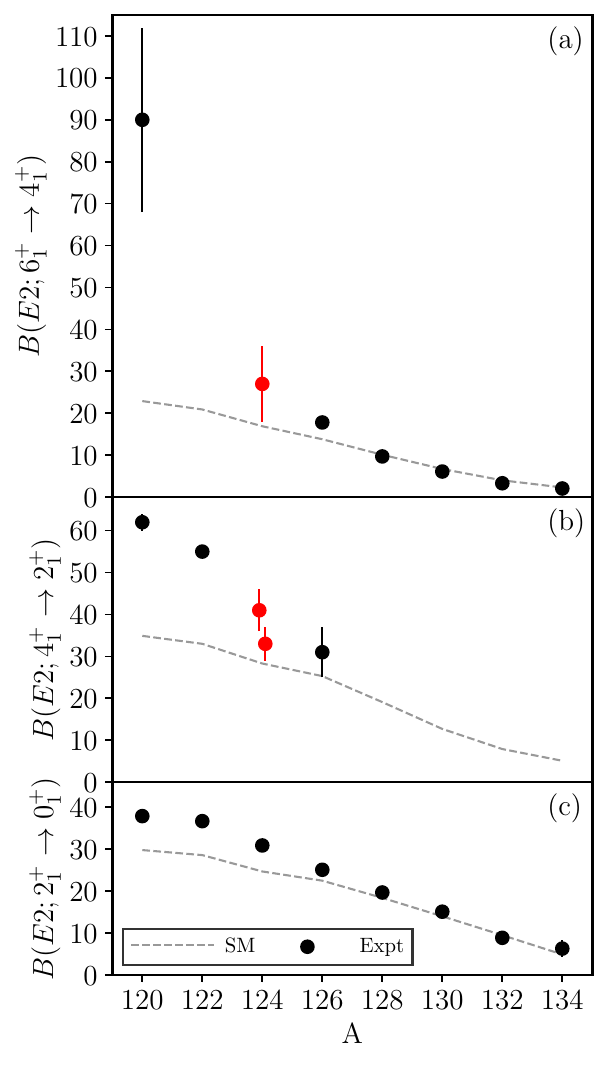}
	\caption{
		Comparison of shell-model $B(E2)$ transition strengths in the Te isotopes with experiment.
		Experimental data shown in black are from Refs.~\cite{Saxena_2014,Hicks_2005,Stokstad_1967,126Te,128Te,130Te,132Te,134Te} and data shown in red are from the present work. The measured $B(E2; 4^+_1\rightarrow2^+_1)$ value was sensitive to the sign combination of the $4^+_2\rightarrow2^+_1$ and $4^+_2\rightarrow2^+_2$ transitions; we have included both values here. \label{fig:8}
	}
\end{figure}

The theoretical and experimental $B(E2)$ values are compared for the $2^+_1 \to 0^+_1$, $4^+_1 \to 2^+_1$, and $6^+_1 \to 4^+_1$ transitions in Fig.~\ref{fig:8}.
Fast-timing measurements of $B(E2;4^+_1 \rightarrow 2^+_1)$ in \nuc{130, 132}{Te} \cite{Kumar_2022} have been omitted since the uncertainties are too large for the data to be useful on this plot.
Recent measurements of $B(E2;4^+_1 \rightarrow 2^+_1)$ in \nuc{128, 130}{Te} \cite{Prill_2022} have also been omitted since the lifetimes are at the limit of the range for the Doppler-shift attenuation method with the $(p,p')$ reaction, and the results disagree with experimental systematics.

Despite the experimental data being incomplete, it appears that theory and experiment agree well for \nuc{128}{Te} to \nuc{134}{Te}, i.e.\ $76 \leq N \leq 82$. For \nuc{126}{Te} and below ($N \leq 74$), the experimental $E2$ strengths rise above the shell-model values, and markedly so for the  $4^+_1 \to 2^+_1$ transitions in \nuc{120,122}{Te}, and the $6^+_1 \to 4^+_1$ transition in \nuc{120}{Te}.
(There are no experimental data for the $6^+_1 \to 4^+_1$ transition in \nuc{122}{Te}, perhaps because the energy of the $6^+_1 \to 4^+_1$ transition is close to that of the $2^+_1 \to 0^+_1$ transition.)
For \nuc{124}{Te}, the shell model accounts for 80(1)\%, 69(7)\% and 63(20)\% of the  $2^+_1 \to 0^+_1$, $4^+_1 \to 2^+_1$, and $6^+_1 \to 4^+_1$ $E2$ transition strengths, respectively.

\begin{figure}[ht]
	\includegraphics[width = 8.6cm]{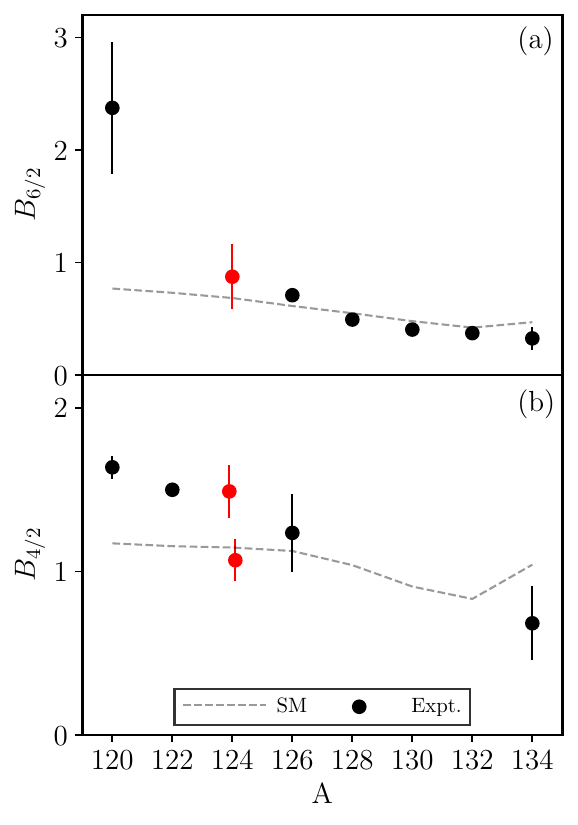}
	\caption{
		Ratios of the (a) $B(E2; 6^+_1 \to 4^+_1)$ and (b) $B(E2; 4^+_1 \to 2^+_1)$ values to the corresponding $B(E2; 2^+_1 \to 0^+_1)$ value in even-even \nuc{120--134}{Te} along with the predictions from the shell model.
		Experimental values for \nuc{124}{Te} (red circles) use $B(E2; 6^+_1 \to 4^+_1)$ and $B(E2; 4^+_1 \to 2^+_1)$ from this work, and $B(E2;2^+_1 \to 0^+_1)$ from Ref.~\cite{Saxena_2014}. The measured $B(E2; 4^+_1\rightarrow2^+_1)$ value was sensitive to the sign combination of the $4^+_2\rightarrow2^+_1$ and $4^+_2\rightarrow2^+_2$ transitions; we have included both values here.
		Experimental values for other isotopes (black circles) are from Refs.~\cite{Saxena_2014,Hicks_2005,Stokstad_1967,126Te,128Te,130Te,132Te,134Te}.\label{fig:9}
	}
\end{figure}

Overall, the shell model gives a good description of \nuc{124--134}{Te}, whereas additional collectivity appears in \nuc{120,122}{Te}.
This view is also supported by the $B(E2)$ ratios, $B_{4/2}$ and $B_{6/2}$, shown in Fig.~\ref{fig:9}.
For \nuc{124}{Te} and heavier isotopes, experiment and the predictions of the shell model remain near the $j^2$ limit for both $B_{4/2}$ and $B_{6/2}$.
That this agreement persists from semimagic \nuc{134}{Te} to \nuc{124}{Te} is intriguing as the wavefunctions certainly become more complex as the number of valence neutron holes increases and the collectivity increases.

According to the shell model, the structures of the $2^+_1$, $4^+_1$ and $6^+_1$ states differ considerably within each isotope and from isotope to isotope in the sequence from \nuc{120}{Te} to \nuc{132}{Te}.
These differences in the structures of the states can be seen by examining the calculated $g$~factors.
Specifically, $g$~factors give insight into how the nucleus carries its angular momentum and particularly how the protons versus neutrons contribute to the wavefunction.
If the states are all part of the $\pi g_{7/2}^2$ multiplet (as in \nuc{134}{Te}), then $g(2^+)=g(4^+)=g(6^+)=g(\pi g_{7/2}) \approx +0.8$~\cite{Stuchbery_2013}. This estimated value of $g(\pi g_{7/2})$ takes $g_l = 1.1$ and $g_s = 0.7 g_s^{\rm free}$; see Ref.~\cite{Stuchbery_2013} and references therein for details of this approach. The large-basis shell-model calculations presented here use the bare orbital $g$~factors, $g_l(\pi) = 1.0$ and $g_l(\nu) = 0.0$, and set $g_s = 0.75 g_s^{\rm free}$ for both protons and neutrons. These are fixed by using the experimental $g$-factor value of the $2^+_1$ state in \nuc{130}{Te}, $g(2^+_1) = 0.351(18)$ \cite{Stuchbery_2007}. For \nuc{124}{Te}, the shell model predicts that $g(6^+_1) = +0.78$, in agreement with the expectations for a $\pi 0 g_{7/2}^2$ excitation. However, the shell-model values of $g(4^+_1) = +0.53$ and $g(2^+)=0.30$ indicate mixed proton and neutron contributions in the 2$^+_1$ and 4$^+_1$ states, more so in the 2$^+_1$ state than the 4$^+_1$ state.
The shell-model $g(2^+_1)$ value is in good agreement with the experimental value of $g(2^+_1) = 0.326(18)$~\cite{Stuchbery_2007}.
Thus, the agreement of the $B(E2)$ ratios with the simple $j^2$ limit is puzzling and likely fortuitous because the $g$-factor ratios are not in agreement with this limit.

Comparison of the experimental and shell-model $B(E2)$ values therefore provides strong evidence that the $6^+_1$ state retains a strong $\pi {(g_{7/2}^2)}_{6^+}$ component, whereas the $2^+_1$ and $4^+_1$ states are mixed proton and neutron excitations.
Measurements of the $g$~factors of the $4^+_1$ and $6^+_1$ states would test this conclusion.

\subsection{Collective versus seniority structure in the Te isotopes}
The present experimental results together with the GCM and shell-model calculations give insights into the emergence of collectivity versus the persistence of seniority structure in the Te isotopes between \nuc{120}{Te} and \nuc{134}{Te}.
Neither model is perfect --- there is an interplay of single-particle (seniority) structure and emerging collectivity as the number of neutron holes increases below \nuc{134}{Te}.

The key observations are that the GCM describes \nuc{120}{Te} quite well, but not \nuc{124}{Te} and the heavier isotopes, whereas the shell model gives a satisfactory description of the $E2$ strengths of \nuc{124--134}{Te} but begins to significantly underestimate the $E2$ strength displayed by \nuc{120}{Te} and \nuc{122}{Te}, particularly for the $4^+_1$ and $6^+_1$ states.
In other words, \nuc{120}{Te} has apparently become weakly collective up to the $6^+_1$ state.

For the isotopes \nuc{124--132}{Te}, the pattern is that collectivity is emerging beginning with the $2^+_1$ state, while the $4^+_1$ and $6^+_1$ states tend to retain a dominant proton seniority structure; this seniority structure persists in the $6^+$ state longer than it does in the $4^+$ state.
Such a pattern has also been recognized in recent studies of \nuc{130}{Te}~\cite{Hicks_2022} and \nuc{132}{Xe}~\cite{Peters_2019}.

The trend of the $E2$ strengths, both in absolute terms and as $B_{4/2}$ and $B_{6/2}$ ratios, suggests a sudden onset of collectivity between \nuc{124}{Te} and \nuc{120}{Te}; however, it should be noted that the $B(E2)$ ratios still fall short of the vibrational model values.

The present $B(E2; 6^+_1 \to 4^+_1)$ result in \nuc{124}{Te}, together with previous data, supports the conjecture of Kerek~\cite{Kerek_1971} and Lee \textit{et al.}~\cite{Lee_1991} that the $6^+_1$ states of the Te isotopes may retain a significant proportion of two-proton seniority structure toward \nuc{120}{Te}.
This interpretation is supported for \nuc{124--134}{Te}, but in the absence of a $B(E2; 6^+_1 \to 4^+_1)$ measurement on \nuc{122}{Te}, it remains unclear whether \nuc{122}{Te} should be grouped with \nuc{124--134}{Te} or with \nuc{120}{Te} (which appears to be collective), or whether it sits at the interface of the transition to low-excitation quadrupole collectivity in the Te isotopes.

\section{CONCLUSIONS}\label{sec:conclusions}
The nucleus \nuc{124}{Te} has been studied by Coulomb excitation with \nuc{16}{O} and \nuc{58}{Ni} beams.
A new target chamber for the CAESAR array, together with an array of photodiodes to serve as particle detectors, was employed to measure particle-$\gamma$ coincidences and deduce $E2$ transition strengths.
The $B(E2; 6_1^+ \to 4_1^+)$ value was measured for the first time.
This transition in \nuc{124}{Te} remains firmly below collective-model predictions and, in line with shell-model calculations, supports the persistence of a two-proton seniority structure for the $6^+$ state.
In contrast, the published data for \nuc{120}{Te} suggest that the $6^+$ state in that nuclide may have become collective.
A higher precision measurement on \nuc{120}{Te} is needed to elucidate the situation.
Furthermore, there are no data on $B(E2; 6_1^+ \to 4_1^+)$ in \nuc{122}{Te}, which appears to be at the point of transition between the persistence of seniority structure evident in \nuc{124}{Te} and the heavier isotopes \nuc{126--132}{Te}, and \nuc{120}{Te}, which appears to have become collective, at least for the low-excitation structure up to the $6^+_1$ state.
Additional experiments are also required to resolve the ambiguities concerning the $2^+_2 \to 2^+_1$ transition strength, mixing ratio and the $2^+_2$-state lifetime.

Despite these uncertainties, it is clear that while collectivity is emerging in the low-excitation spectra of the Te isotopes below the $N = 82$ isotope \nuc{134}{Te}, the two-proton seniority structure, $\pi 0g_{7/2}^2$, persists in the $6^+_1$ state to at least \nuc{124}{Te}.
Additional systematic measurements are needed to fully expose the structure of the Te isotopes.
In particular measurements of $E2$ and $E0$ transition strengths, as well as $g$~factors, are required.

\begin{acknowledgments}
	The authors wish to acknowledge the excellent work of the Technical Staff of the ANU Department of Nuclear Physics and Accelerator Applications.
	We thank T.~Kib\'edi for valuable discussions about uncertainties, J.~Heighway for target production, and T.~Tunningley and R.~Tranter for their work on the target chamber and detector mounts.
	We also thank H.~A.~Alshammari, J.~T.~H.~Dowie, A.~A.~Gopakumar, T.~Kib\'edi, R.~du~Rietz, B.~Swinton-Bland, B.~Vallette, and Y.~Y.~Zhong for their contribution to data collection.
 	Figure~\ref{fig:1} in this article was created using the LevelScheme scientific figure preparation system~\cite{Caprio_2005}.
	This work was supported in part by the Australian Research Council Grant No. DP210101201 and the International Technology Center Pacific (ITC-PAC) under Contract No. FA520919PA138.
	M.~R.\ gratefully acknowledges support from the ANU Research School of Physics.
	L.~J.~M.\ acknowledges support of the Australian Government Research Training Program Scholarship.
	V.~U.~B.\ and N.~J.~S.\ acknowledge stipend support from the Australian Research Council Centre of Excellence for Dark Matter Particle Physics.
The authors acknowledge the facilities, and the scientific and technical assistance provided by Heavy Ion Accelerators (HIA). HIA is supported by the Australian Government through the National Collaborative Research Infrastructure Strategy (NCRIS) program. 
 \end{acknowledgments}



%


\end{document}